\documentclass[12pt]{article}

\usepackage{graphicx}
\usepackage{epsfig}
\usepackage{amsmath}
\newcommand{\be}{\begin{equation}}
\newcommand{\ee}{\end{equation}}
\newcommand{\bea}{\begin{eqnarray}}
\newcommand{\eea}{\end{eqnarray}}
\newcommand{\ba}{\begin{array}}
\newcommand{\ea}{\end{array}}

\begin{document}

\begin{center}
{\large \bf Energy conversion in isothermal nonlinear irreversible processes - struggling for higher efficiency
\footnote{On the occasion of the 60th birthday of Ulrike Feudel}}\\
\vspace*{7mm}
{\footnotesize Werner Ebeling, Rainer Feistel}\\
\vspace*{3mm}
   {\small\em Institute of Physics, Humboldt-University Berlin,}\\
	{\small\em Leibniz Institute for Baltic Sea Research, Warnem\"unde, Germany}\\
{\small(\em{to be published in Eur. Phys. J. ST 2017)}}
\end{center}


\begin{abstract}
First we discuss some early work of Ulrike Feudel on structure formation in nonlinear reactions including ions
and the efficiency of the conversion of chemical into electrical energy. Then we give some survey about
energy conversion from chemical to
higher forms of energy like mechanical, electrical and ecological energy. We consider examples of energy conversion in several natural processes and in some devices like fuel cells. Further,  as an example,  we study analytically the dynamics and efficiency
of a simple "active circuit" converting chemical into electrical energy and driving
currents which is roughly modeling fuel cells.
Finally we investigate an analogous ecological system of Lotka - Volterra type consisting of an "active species" consuming some
passive "chemical food". We show  analytically for both these models that the efficiency increases with the load, reaches values higher then 50 percent in a narrow regime of optimal load and goes beyond some maximal load abrupt
to zero.
\end{abstract}

\section{Introduction and reminiscence}

Just after the appointment of one of the presents authors as a Professor at Humboldt University Berlin in 1979, a young energetic student entered his office. Let us try to repeat briefly the conversation: "I am Ulrike Wauer and would like to work on selforganization in your group".
The reply was like: "What is your specific interest and by the way, are you related to the organist Hans - G\"unther Wauer",
and her answer was like:
"My interest are problems related to biophysics, and yes, I am his daughter".\\
This happened just in the time when the work of Turing (1952) \cite{Turing52} and the papers of Prigogine and Nicolis since 1967
\cite{PriNic67,NicPri77}  on models of structure formation
in chemical reactions became popular in  the scientific community \cite{Eb76}.
The subsequent  discussions included the second author and led to the conclusion that it could be of interest to work on the
role of ionic components in nonlinear reactions, to study chemical and electrical structures and
the efficiency of energy conversion and to look for biophysical applications \cite{EbeFeu83,FeuFeEb84FeuEb84}.\\
As well known, in nearly all biophysical processes ions are participating and further in all irreversible processes energy is
changing its form and entropy is created in the interior.
Since Clausius the efficiency defined as the relation of useful energy output to energy input is the crucial problem.
For reversible and almost reversible processes, this is a well studied topic, for nonlinear processes far from equilibrium
there are several studies following the early works \cite{Ross,Feistel1985,FeEb89,Lipowsky,FeEb2011}.\\
Anyhow the problems in this field seem to be difficult and in part still controversial.\\
Ulrike Wauer - Feudel started her career with two papers about the efficiency of dissipative structures including
ionic reactions \cite{EbeFeu83,FeuFeEb84,FeuEb84}. Everything worked out fine, chemical inhomogeneities, electrical structures, self - electrophoresis and driven currents were found.
 For example, a membrane - coupled 2-box chemical battery permitted an analytical solution for the bifurcation diagram with various transitions between stable and unstable nodes and foci \cite{FeuFeEb84}. The response of the system to a load which was modeled by the resistance of an external conductor
 was studied and the efficiency of the energy conversion was analytically calculated. The efficiency was defined as the relation of that part of input chemical energy converted in the external conductor considered as the  working load. A thesis was written and successfully defended. However, one point remained disappointing:
In several examples based on more or less random choice of the chemical parameters, the efficiencies of the conversion of chemical into electrical energy occurred to be rather low, around $10^{-4}$
\cite{FeuFeEb84,FeEb89}. Therefore we concluded that the idea was good, however, the numbers we obtained did not support the
hope for interesting practical applications. However, we considered just special parameter sets and a systematic search for the global optimum of efficiency in the space of the many free parameters of the "generator" was still missing.\\
In the present work we plan to take up the problem again and to show:\\
1. There is no severe theoretical bound limiting the isothermal energy conversion in chemical reactions from above
like the Carnot bounds for thermal machines, so the problem remains to find examples with good efficiency \cite{Lipowsky,FeEb2011}.\\
2. Energy conversion from chemical sources is always dependent on nonlinearities connected with
many parameters, therefore the optimization of efficiency is a multi - parameter problem,
"trial and error" or "searching in the fog" are not promising strategies \cite{PPSN}.\\
3. Checking for the role of specific nonlinear effects, in particular those connected with electrode processes are of relevance. The careful study of simple and solvable optimization problems may be helpful
to develop strategies to find "good efficiencies", what may mean better than 50 percent.\\
 Recent research explored one of the basic mechanisms of life connected with the conversion of chemical into mechanical energy.
From precise experiments we got a lot of insight how in the cells on the nano level the available chemical energy is stored in ATP molecules and converted to mechanical motion and electrical signals.
In particular we know from experiments details about the functioning and efficiency of single molecular motors. This is one of the hot topics of recent research. There are e.g. new experimental findings that the important biophysical mechanism in cells based on the F1-motor may achieve rather high efficiency of energy conversion \cite{Toyabe13}.
In typical experiments with simple organisms
converting ATP-energy into motion, it has been observed that as a rule the efficiency increases first with the load, reaches quite high values and then it decreases and goes to zero \cite{Nishiyama,Harada}.\\
Quite generally, living organisms represent complex dissipative structures with rather high efficiency of converting imported chemical
energy to mechanical or electrical work \cite{Reichholf2011}. This sometimes high efficiency remains a
challenging recent target for scientists and engineers, for example for the buffering of technically produced "green" energy in
chemical batteries with minimum input-output losses \cite{Feistel1985,FeEb89,FeEb2011}.\\
In the present work we start with a brief survey about efficiencies of the conversion of chemical to electrical energy. Then we
are going to analyze a simple electrical dissipative structure consisting of a circuit modeling an electrochemical cell, which is mathematically related to a recent study of the efficiency of molecular motors \cite{Eb15}. We identify an essentially autocatalytic term in the current
which is responsible for the formation of an electrochemical dissipative structure.
Then we transfer the model which was originally developed for a mechanical motor to a dissipative ecological structure.
We use here the analogy of ecological systems of Lotka - Volterra type to Hamilton systems generalizing
the approach developed in \cite{FeEb89}.
We include the discussion of ecological problems just for showing possibly relations to more recent work of the group of Ulrike Feudel
on ecological dynamics and control of nonlinear systems \cite{Dutta,Pisar}.\\

\section{Survey for efficiency of the energy conversion from chemical to higher forms of energy}

\subsection{Several mechanisms of energy conversion by means of irreversible processes }
Conversion of energy by means of irreversible processes and dissipative structures is not new; various technical and biological systems function this way. If we restrict ourselves to systems with chemical energy input, we may distinguish between converters to, e.g., mechanical, electrical or radiation energy. Here we briefly discuss typical efficiencies of some traditional such "engines".
Mechanical work is obtained from the chemical energy of fuels such as wood, coal or oil by self-sustained nonlinear oscillators such as steam engines, combustion motors or jet plane turbines. As far as the fuel is burned and the resulting heat is used as an energy pump of the nonlinear oscillator, the efficiency of such machines cannot exceed that of the thermodynamic Carnot cycle, as a consequence of the 2nd law.
Steam engines have recently returned to the focus of engineering in the context of solar power plants. In 1698, Thomas Savery patented a commercial steam engine as "a new invention for raising of water and occasioning motion to all sorts of mill work by the impellent force of fire. Without pistons, the machine could lift water to a maximum height of 9 m. Already in 1690, as a Huguenot refugee in Germany, Denis Papin had built the very first piston steam engine as a model. Inspired by Savery's invention and with the help of Gottfried Leibniz, he constructed a functioning steam engine in 1705. However, it was James Watt's famous steam engine of 1781 that eventually revolutionized the factories of Great Britain and initiated the global industrial age. His seminal invention is honored by a monument in Londons Westminster Abbey, along with Isaac Newton and Charles Darwin. The efficiency of traditional steam engines did not exceed 10 percent. This value could be raised to almost 20 percent by technical means such as increased temperatures or pressures, and by advanced management to almost 50 percent in modern electrical power stations.
A candle flame is a dissipative structure driven by the chemical energy of wax. Depending on its boundary conditions, it may work in steady-state or oscillating regimes. With respect to the candles radiation energy output, a typical efficiency may be 17 percent \cite{Hamins2005}. The kinetic energy of the produced air flow is the target of stronger flame versions that are burning in gas turbines, such as aircraft propulsion engines, or in rockets. Combustion turbines are characterized by bifurcation diagrams \cite{Soares2014} that may be similar to those of candles. The worlds first jet plane started in 1939 from Marienehe near Rostock in Germany \cite{Warsitz2009}. As a part of the overall efficiency, the thermodynamic cycle efficiency is typically about 60 percent in rocket engines and 30 percent in turbojets.

\subsection{Energy conversion cascades in several natural phenomena}
The processes of selforganization and evolution on our Earth are based on a big energy conversion machine driven by a flow of sunlight
from the sun to the Earth and back to the empty space around the Earth. The corresponding entropy export was estimated to have the amount of
$ 1 W / K$ per square meter of the surface of the Earth \cite{FeEb89,FeEb2011}.\\
In order to give more special examples, the analysis of hurricanes as compared to a Carnot process was analyzed \cite{Emanuel1991}.
Recently also the efficiency of the atmospheric hydrological cycle was estimated \cite{Laliberte2015}.
As an example of self-organized criticality, for the work released by the Chilean 2010 earthquake at Concepcion, a
very small efficiency smaller $10^{-6}$ percent
was estimated in \cite{FeEb2011}.\\
In some recent book we find estimates of the efficiencies of the food cycles driven by the sunlight \cite{Balmer}.
According to this estimate the efficiency of of the conversion of sunlight to plant growth is only 1 percent and the efficiency of plant eaters by grazing
herbivores is 20 percent.  However the efficiency of the carnivores who hunt and eat herbivores is only around 5 percent, so that the total efficiency
of the cycle is not higher than $10^{-2}$. Therefore we should be careful with conclusions about the high efficiency of processes in nature \cite{Reichholf2011}.
Evolution is optimizing for high efficiencies only in special cases where the energy sources are rare in the input flows are small.
In most cases simple functioning of a mechanism is more
important than efficiency of energy conversion. Therefore our special interest is targeted to those
examples where the efficiency of energy conversion is indeed high.

\subsection{Energy conversion in muscles}
All higher forms of life are connected with active motion  based on muscles.
The fundamental energy transduction process in muscles
is the conversion of chemical free energy (from ATP
splitting) into mechanical energy by myosin crossbridges.
The fraction of the available free energy converted into
mechanical work is called the thermodynamic efficiency $\eta$ . The reported maximum efficiency values for the human
muscle are high compared to the value expected on the
basis of efficiency values obtained in experiments using
isolated muscles from other species. For example, Barclay et al. \cite{Barclay}
report for human muscles efficiencies between 40 and 70 percent.
Reported values of efficiencies for non-mammalian muscle are
generally lower than those described above for human
muscle. In studies using isolated, non-human mammalian
muscles, the fraction of free energy converted into work by
cross bridges, under conditions designed to maximize
efficiency, is typically between 20 and 30 percent.

\subsection{Molecular motors}

Molecular motors, we speak here about motor proteins, are enzymatic
molecules that convert chemical energy, typically obtained from the
hydrolysis of ATP, into mechanical work
and motion \cite{Lipovsky}. Several motor proteins, such as kinesin, dynein, and
certain myosins, step unidirectionally along linear tracks, specifically
microtubules and actin filaments. These molecular motors play a crucial role in
cellular transport processes, organization, and function \cite{KoloFisher}.
In particular in connection with the investigations on molecular motors, recent results on
active Brownian motion driven by a chemical energy depot are of interest \cite{Romanczuk12}.
In this connection a quite simple model has been developed for the conversion of chemical energy to mechanical motion, the depot model (sometimes called SET-model) which is based on the assumption of a depot filled with
chemical energy coupled to a mechanical system. In addition to Langevin type equations for position and velocity
of the particles a balance equation for the flow between the source and an internal energy
depot was introduced \cite{Romanczuk12,ScEbTi98,EbScTi99,EbGuFi08,ErEbScSc00,SchwBook,Tilch99}.
The existence of a depot with an internal variable is what this class of models differs from other models
of active motion \cite{Romanczuk12}. We mention several applications of the depot model to cell motility \cite{BiSc11,GaBiSc11}, and to modeling molecular motors \cite{Romanczuk12,RoKaEb13}.
We will show here that some formerly found lower limits of efficiency \cite{Romanczuk12,RoKaEb13,Zabicki,Zabicki2,FiEbGu13} can be lifted under special conditions if nonlinear effects are included.
In a narrow range of optimal load and in case of very low loss the SET model motor may come close to 100 percent efficiency. \\
For the given reasons and in particular looking at the new experimental results we believe that it may be timely to take up the problem raised by Ulrike Feudel again and to analyze it  in the light of recent experimental and theoretical results.\\

\subsection{Energy conversion in fuel cells}

\begin{figure}[htbp]
 \begin{center}
  \includegraphics[width=12cm]{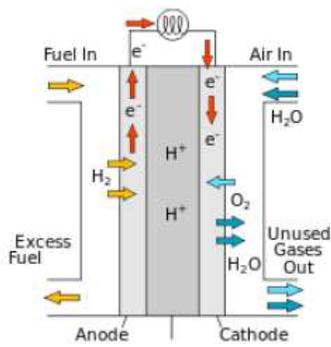}
  \caption{Schematic operation of a fuel cell converting the energy of hydrogen gas into electrical energy. Note that the essential "autocatalytic" nonlinearity is connected with the processes at the electrode}
 \end{center}
  \label{fuelcell0}
 \end{figure}

Fuel cells are devices that convert chemical energy to electrical energy without the intermediate stage of producing and consuming heat. Typical efficiencies of fuel cells range from 20 percent to 70 percent \cite{Bachmann2013}. Similar to the dissipative structures of steam engines or gas turbines, a fuel cell will "starve" at subcritical values of the energy pumping rate. In this article, a theoretical model will be described for alternative, self-organized processes that may act like a fuel cell.
As already said, a fuel cell is a device that converts the chemical energy from a fuel like hydrogen gas into electricity through a chemical reaction of positively charged hydrogen ions with oxygen or another oxidizing agent. This process requires a continuous source of fuel such as hydrogen and oxygen to sustain the chemical reaction of the fuel with the oxidizing agent. Fuel cells can produce electricity,
by exploiting the input of chemical energy, as schematically shown in Fig. \ref{fuelcell0}.

For a fuel cell to produce electricity, it must be continually supplied with fuel and oxidant. The details of this mechanism can be quite complicated.
We will study here only a simplified schema which we call {\it "active circuit"}. The model should reflect at least in a rudimentary form the following observed physical properties of fuel cells:
When a fuel cell is operated with high output, i.e. high current, its demand for reactants is large. If the reactants are not supplied to the fuel cell quickly enough, the device will starve.
The work of fuel cells is based on electrochemical reactions, connected with the fuel, e.g. hydrogen, the oxidizer and charges.
Once the reactants are delivered to electrodes, they must undergo electrochemical reactions. The current generated by the fuel
cell is directly related to how fast the electrochemical reactions proceed. Fast electrochemical reactions result in a high current output from the fuel cell.
Slow reaction results in a low current output.
Ionic conduction happens through the electrolyte and electron conduction through the external circuit.
The electrochemical reactions occurring in step 2 either produce or consume ions and electrons. Ions produced at one electrode must be consumed at the other electrode. The same holds for electrons. To maintain charge balance, these ions and electrons must therefore be transported from the locations where they are generated to the locations where they are consumed. For electrons this transport process is rather easy. As long as an electrically conductive path exists, the electrons will be able to flow from one electrode to the other.\\
Our (simplified) schema to model these quite complex processes is the following:\\
1. The active particles in our schema are the charges. For simplicity we study only two variables for the characterization of the electric state, the charge of the capacitor $Q$
and the current ${\it I (t)}$ in the circuit. The charge is a  variable which may increase or decrease and is therefore a dynamic quantity $Q(t)$.
The charge may increase or decrease following the consumption of fuel and is connected with a current ${\it I (t)}$. The third relevant dynamical variable is the energy contained in the chemical reactor $E(t)$ which is driving the electric current ${\it I}$ in the circuit.
This way we have to define 3 differential equations that specify
the dynamics of the fuel cell within our model
\bea
\frac{d Q(t)}{d t} = {\it I (t)},\\
\frac{d {\it I(t)}}{d t} = F_2 ({\it I (t)}, E(t), Q(t)),\\
\frac{d E(t)}{d t} = F_3 ( E(t), {\it I (t)}).
\eea
The two functions $F_2, F_3$ modeling the energy flow from the chemical reactor / capacitor to the current in the circuit have still to be determined in a way that
the properties of the fuel cell, described above, are reflected at least in a qualitative way. This will be done in the next section.\\
There are several properties which are relevant for an efficient regime of fuel cells, which however are not principal ingredients and will not be modeled here. For example, the efficient delivery of reactants is more effectively accomplished by using flow field plates in combination with porous electrode structures. Flow field plates contain many fine channels or grooves to carry the gas flow and distributes it over the surface of the fuel cell. The shape, size, and pattern of flow channels can significantly affect the performance of the fuel cell.
Obviously, high current output is desirable. Therefore, catalysts are generally used to increase the speed and efficiency of the electrochemical reactions. Fuel cell performance critically depends on choosing the right catalyst and carefully designing the reaction zones.

\section{Circuit under load driven by an electrochemical cell}
\subsection{The model of active circuit for a fuel cell }
In previous work we studied active Brownian particles. This model considers mechanical motion including coupling to an energy depot. We considered a mass $m$, located at coordinate $q(t)$  moving with velocity $v(t)$ where $v = \dot q$,
which is accelerated at the cost of the chemical depot energy $e$ and may perform work against a load:
\begin{equation}
\label{Mdyn}
  m {\dot v} + \rho_0 v + \kappa q =  F_0  + d e v,  \qquad F_0 = - a, \qquad a < 0.
\end{equation}

Here $F_0 <0$ is an external force, a load, and for the driving force on cost of a depot energy we assumed
$F_d = d e v$. This driving force is proportional to the energy depot of chemical nature
$e$ , $\rho_0 = m \gamma_0$ a friction constant, and $\kappa$ some elastic constant, which in most cases is put to zero here.\\
Within the original model developed by Schweitzer - Ebeling - Tilch (sometimes called SET model) we assumed for the
depot energy $e(t)$ the simple dynamics
\cite{Romanczuk12,ScEbTi98,ErEbScSc00,RoKaEb13}.
\begin{equation}
\label{depotdyn0}
  \dot{e} =  q_c - c e  - d {\bf v}^2 e.
\end{equation}
With respect to its mathematical form this model belongs to the class of systems studied already in \cite{FeEb89}.
The 3 constants $q_c$, which is the input rate of depot energy, $c$, the decay rate and $d$, the rate of transmission of depot energy to energy of motion, determine the functioning of the motor mechanism. We considered $e$ as a kind of
chemical energy as e.g. ATP stored in the depot \cite{RoKaEb13}.
The physical meaning is that we have a permanent inflow of
energy, which is constant $q_c =const$, and
flows with rate $d v^2(t) e(t)$ to the mechanical
degree of freedom \cite{ScEbTi98,EbScTi99}. Note
that previously we mainly used units leading to $m=1$.
For $q_0 > \rho_0 c / d$ there exists a stationary point in the positive cone of the energies which corresponds to two stationary points in the phase space.
This corresponds for the force-free case to two velocities for the same stationary depot energy
\begin{equation}
v = \pm v_0, \qquad e = e_0 = \rho_0 / d, \qquad  v_0 = \pm \sqrt{\frac{q_c}{\rho_0} - \frac{c}{d}}.
\end{equation}
We note that the coordinate $q(t)$ is not necessarily a linear length coordinate but may instead be the angle coordinate of a rotor  \cite{Romanczuk12,RoKaEb13}.\\
We will show now that the SET model may be mapped to a simplified model of a fuel cell. In order to proceed we use a special version of the
SET - model in a form which considers only the energy balance for the kinetic energy $k = m v^2 / 2$ which we obtain by multiplying
the equation of motion with the velocity
\bea
\frac{d k}{d t} + 2 \frac{\rho_0}{m} k = F_0 \cdot v + 2 d e k\\
\frac{d e}{d t} + c e = q_0 - ce  - 2 \frac{d}{m} e k
\eea
This version of the SET - model may be more easily converted to our schema for the operation of a fuel cell which we name "{\it active circuit}".
We will use the (formal) analogy between mechanical and electrical systems by using the "dictionary" \cite{FeEb89}:\\
mechanical coordinate $q , \qquad$ - electrical charge $Q$,\\
velocity $v = {\dot q} \qquad$ - current $\it I$,\\
kinetic energy $\sim v^2 \qquad$ - current energy $\sim {\it I}^2$ , \\
mechanical mass $m  \qquad$ - impedance $L$,  \\
oscillator constant $\kappa  \qquad$ - reciprocal capacity $1 / C$ , \\
friction $\rho = m \gamma_0 \qquad$ - resistance $R$, \\
driving force $F_d \qquad$ - electromotoric force ${\it U}_d$. \\

The electrical analog to a mechanical system is an R - C - L - circuit and the dynamics is according to our dictionary
\bea
\label{Idyn}
    \frac{d Q}{d t} = {\it I}; \\
     L \frac{d \it I}{d t} = - R {\it I} - \frac{Q}{C} +  {\it U}_0  + d E {\it I}^2;\\
     \frac{d E}{ dt} = q_c - c E - d E {\it I}^2.
\eea
Here ${\it U}_0 = - a < 0$ is a kind of external load which is to overcome, let us assume that a charge $Q$ is to transfer
to a higher level of the electrical potential. Admittedly, this model is oversimplified. Usually, the equivalent circuit for an electrode involves  a parallel connection
of a capacitance and a faradaic impedance, the total current thus being composed of a capacitive contribution and a contribution due to the electrochemical reaction. We concentrated here on modeling the energetic aspect, i.e. the physical fact that the chemical energy generated by the reaction is converted to the energy of the electrical circuit
and that this process is self - enhancing.
The reason for neglecting further effects here is not a physical one. Our motivation is, to allow a one to one mapping to the solvable mathematics of the model SET. Any additional term would break he possibility for analytical solutions.
Our model is in fact nothing more than the extension of the standard circuit equations by a quadratic term in the current which stands for the energy input. This term is, observing the energy balance, coupled to a simple equation for one reaction.\\
We consider this as a minimum of relations describing the essence of the energy transfer reaction - current.\\
The properties of this dynamical system have been studied for the mechanical picture in the works \cite{EbScTi99,Tilch99}. Here we concentrate on  the
stationary states and questions of energy conversion. \\
For the case of no extra voltage  $U_0=0$ and large enough chemical input, there exists always a stationary state
\be
E_0 = \frac{R}{d}; \qquad I_0^2 = \frac{q_c}{R} - \frac{c}{d}
\ee
In this state a stationary current is flowing in the circuit which is sustained by the inflow of chemical energy
and is dissipated in the resistance $R$.\\
The dynamics of the energy exchange between circuit and fuel cell is described by the balance equations for (kinetic) electrical energy of the current $E_e \sim {\it I}^2$
and the  energy of the chemical reactor  $E(t)$.
\bea
\frac{d E_e}{d t} + 2 \frac{\rho_0}{m} E_e = U_0 \cdot {\it I} + 2 d E  E_e,\\
{\dot E} = q_c - c E - \frac{2 d}{L} E E_e
\eea
which may be written as an equation for the current
\bea
L \frac{d {\it I}^2}{d t} + R {\it I}^2 = U_0 \cdot {\it I} + d E {\it I}^2\\
{\dot E} = q_c - c E -  d E {\it I}^2.
\eea
Here $q_c$ denotes again the inflow of chemical energy, $c E$ denotes internal loss (internal resistivity)
and the last term denotes the energy transmitted per unit time to the R-L-circuit which has to appear there
as a positive energy support ${\it U}_d = + d {\it I}^2 E$ .
This is in fact an auto - catalytical term which is essential for the functioning of chemical dissipative structures. Let us still note that in the stationary state the value of the impedance $L$ does not play any role
for the work of the cell.

\subsection{Efficiency the conversion of chemical to electrical energy in a fuel cell}

\begin{figure}[htbp]
 \begin{center}
  \includegraphics[width=8cm]{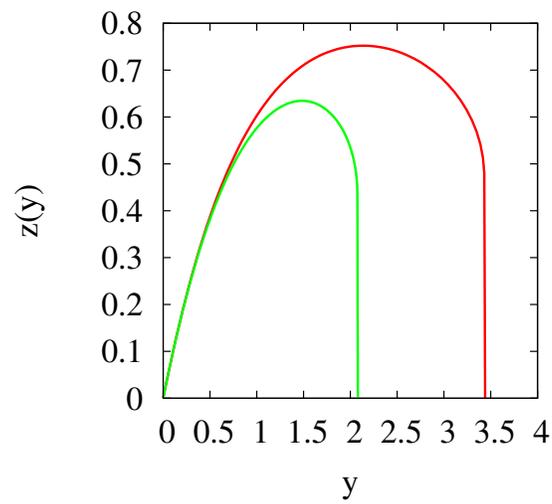}
  \linespread{1.2} 
  \caption{Active circuit under conservative load: Efficiency $z$ vs. conservative load $y$ for two quite low circuit loss parameters $\lambda = 0.05, 0.02$.
  For finite loss parameter values, the
  curve stops at some critical load, corresponding to regions where the cubic equation has no real solutions.
  Then the working regime of the circuit breaks suddenly down.}
  \label{Zvsy}
 \end{center}%
\end{figure}

We study now the efficiency, which is as usual defined as the relation of energy flow used for some purpose to the
total imported energy flow. There are two situations of interest:\\
1. The  energy is used to overcome an additional external dissipation, an extra resistance,
and is converted to heat (or possibly light).\\
2. The energy is used to do useful mechanical or electrical work, possibly by bringing a charge to a higher
potential level or driving an electromotor.\\
The first case is much easier in mathematical respect. We model the external dissipation by some additional unsymmetrical resistor $R_1$.
The model assumption is that the resistor which simulates a load acts only on positive currents, but not on negative currents.
Then the negative currents remain unchanged however the positively - directed currents will go down by some amount
${\it I}_1 < {\it I_0}$ determined by
\be
{\it I}_1^2 = \frac{q_c}{R + R_1} - \frac{c}{d}
\ee
In other words the shifted attractors are
\be
{\it I}_1 = + \sqrt{\frac{q_c}{R + R_1} - \frac{c}{d}} \qquad if \qquad {\it I} > 0
\ee
and
\be
{\it I}_1 = - \sqrt{\frac{q_c}{R} - \frac{c}{d}} \qquad if \qquad {\it I} \le 0
\ee
The additional loss to overcome by the  positive currents is $R_1 {\it I}_1^2$. This way we find for the efficiency
\be
\eta = \frac{R_1 {\it I}_1^2}{q_c} = R_1 \Big(\frac{1}{R + R_1} - \frac{d}{c}\Big)
\ee
\begin{figure}[htbp]
 \begin{center}
  \includegraphics[width=5cm]{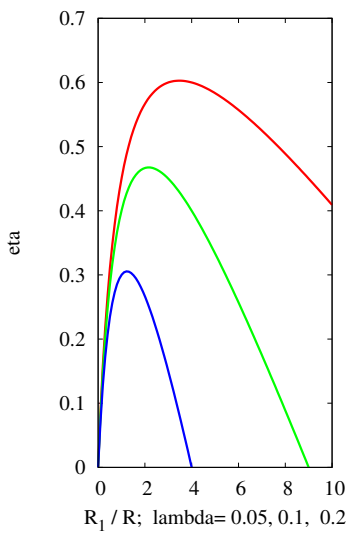}
  \caption{Active circuit under dissipative load: Efficiency of energy conversion $\eta$ vs. relative load $R_1/R$. The system parameters are $\lambda = (R d / c q_c) = 0.05, 0.1, 0.2$. The efficiency increases first with the load and reaches a maximum around $R_1 / R \approx 1 - 4$. Beyond a critical relative load, the system stops to work. }
 \end{center}
  \label{effi1}
 \end{figure}
Now we investigate the case of constant counter voltage modeling a load. This is the situation where it is more difficult to bring the charge to a higher potential level. This situation is physically quite simple but mathematically more difficult. Without loss of generality
we assume that the additional voltage is directed to below $U_0 = - a, \qquad a > 0$ i.e. it tends to decrease the current.
If $a$ is small, the two attractors are shifted linear in the force to lower values of the current
\begin{equation}
{\it I}_1 = \pm {\it I}_0 - \frac{q_c a}{2 R^2 {\it I}_0^2}; \qquad E_1 = \frac{R}{d} \pm \frac{a}{{\it I}_0}
\end{equation}
For larger voltages, we have to solve nonlinear equations which we formulate in dimensionless variables:
$$
y = \frac{a}{R {\it I_0}} = \frac{a}{R {\it I}_0}; \qquad \xi  = \frac{{\it I}_1}{{\it I}_0}
$$
We find a cubic equation which is an implicit relation between current and load:
\begin{equation}
\label{cubic}
\xi^3 + y \xi^2 - \xi + \lambda y = 0; \qquad \lambda = \frac{c \gamma_0}{q_c d -c \gamma_0}.
\end{equation}
Here $\lambda$ is a parameter describing the dissipation in the circuit and $\lambda y$ determines the character
of the solutions. In the context of stochastic transitions the case  is of interest that the positive stationary currents
approach each other and merge finally. The existence of real roots is bound to the condition that the discriminant of
the cubic equation is positive $D(y,\lambda) > 0$. For $D (y , \lambda) < 0$ only downhill (i.e. less interesting) solutions exist. Looking for example at motor parameter $\lambda = 0.5$ the largest $y$ which
provides positive solutions for $\xi$ is $y_{max} =0.28$.
Our driven system is able to do work at the cost of chemical
energy imported by the energy input $q_c > 0$. We define the thermodynamic efficiency as useful work against input of (chemical) energy into the depot
\be
\eta = \frac{|U_0 {\it I}_1|}{q_c} = \frac{a {\it I}_1}{q_c}.
\ee
Here ${\it I}_1 = \xi_1 {\it I}_0 $ is the stable "uphill" current corresponding to the largest positive root $\xi_1 > 0$.
In some previous work \cite{Romanczuk12,RoKaEb13} we used a linear approximation and found a parabolic dependence between efficiency and load with the maximum of efficiency
\be
\eta_{max} = \frac{1}{2} (1 - \lambda)^2.
\ee
which cannot exceed 50 percent reached for the case of no losses. Including nonlinear effects we find using the dimensionless variables $y, \xi$  for the efficiency $\eta = (1 + \lambda) \xi y$
the following equation for the variable $z = \xi y$
\be
z^3 + y^2 z (z-1) + \lambda y^4 = 0
\ee
The solution expressing $y$ in term of the efficiency variable $z$ reads
\be
y^2 = \frac{1}{2 \lambda} z(1-z) \pm \sqrt{\frac{1}{4\lambda^2} z^2 (1-z)^2 - \frac{1}{\lambda} z^3}
\label{yvsz}
\ee
The condition that the root in eq.(\ref{yvsz})should be real gives the largest possible value of efficiency
\be
\eta_{max} = (1 + \lambda) z_{max}; \qquad z_{max} = 1 + 2 \lambda - 2 \sqrt{\lambda(1 + \lambda)}
\ee
The optimal load is the value where the maximum is reached
\be
y_{opt}^2 = \frac{1}{2 \lambda} z_{max} (1- z_{max}) = \frac{1}{\lambda^{1/2}}z_{max}^{3/2}
\ee
In the limit of small losses, the efficiency converges to 1 and the optimal load diverges as $\lambda^{-1/4}$.
To give some examples, if $\lambda = 0.5$ the largest value of efficiency is $\eta_{max} = 0.18$ and for $\lambda = 0.02$ we get the largest value $\eta_{max} = 0.74$ at $y_{opt}=2.1$. A graphical representation of the efficiency variable $z$ against the load $y$ is shown for $\lambda = 0.02$ and for $\lambda =  0.05$ in Fig. \ref{Zvsy}.

Analyzing Fig. \ref{Zvsy} shows that according to nonlinear effects the efficiency curve is not parabolic as the linear theory predicts \cite{Romanczuk12,RoKaEb13} but less symmetric and the maximum can be higher. It is interesting to note that in our case
only nonlinear effects provide efficiencies better then 50 percent. For low losses
and optimal load the efficiency may approach one. However, for any finite quality parameter $\lambda$, the curve stops at some critical load, which corresponds to parameter regions where the cubic equation has  no more real solutions. Then the uphill regime breaks down suddenly, not gradually. Only for the case of no losses $\lambda = 0$ the curve may reach for a special load the efficiency one. As the graphical solutions for the case of very low losses $\lambda = 0.05, 0.02$ clearly demonstrates, the efficiency may only for such low losses and optimal load reach values exceeding 60 or even 70 percent. In order to reach an efficiency of 100 percent, the internal losses in depot and motor must disappear $c \gamma_0 \rightarrow 0$.
In several simulations of the dynamical system we observed for larger loads, say,
$y > 1$ a low stability of the
"hill-climbing state" in the numerical simulations. This means, that it might be quite difficult to
realize the states with high efficiency. In order to overcome this problem we
investigated also the influence of noise and the influence of perturbations
of the external force \cite{Eb15}.\\
The easiest way to estimate the influence of noise is based on the property that our dynamical systems has a
quasi - Hamiltonian character and a canonical distribution function with the probabilities \cite{FeEb89}
\begin{equation}
P \sim exp[- \beta H_{\mathrm{eff}}]
\label{canon}
\end{equation}
where $\beta$ is the inverse noise temperature and $H_{eff}$ is an effective Hamiltonian.

\section{Ecological systems of Lotka - Volterra type driven by food input}
\subsection{Modeling an ecological energy conversion}
We introduce now an ecological analogy which is equivalent from the dynamical point of view.
Formally we get the ecological analogy of the mechanical-electrical circuit
but using the known analogy between Hamiltonian and Lotka - Volterra systems \cite{FeEb89}.
We start from the conservative version of the mechanical system eq.(\ref{Mdyn}) and eq.(\ref{depotdyn0}) and introduce the variables $ X = m v^2 /2, Y = e$ \cite{RoKaEb13,Eb15}. Then
we get for the case that there is no external load $a = 0$ after introducing new constants
$q_c \rightarrow Q_0$, $2 d /m \rightarrow a$ and $2 \gamma_0 \rightarrow b$ the Lotka-Volterra-like version of our
dynamic equations:
\begin{equation}
\label{Lotka1}
  {\dot X} = a X Y - b X; \qquad {\dot Y} = Q_0 - c Y - a X Y
\end{equation}
This version of the dynamical equations has an ecological interpretation.
Following Lotka and Volterra, in an ecological interpretation the $X$ - variable denotes the quantity of a predator - like species and $Y$- the quantity of a prey - like species living from some constant food support $Q_0$. This conservative dynamical system has an ecological meaning only in the positive cone $X, Y \le 0$
and possesses for
$Q_0 > b c / a$ a stationary state in the positive cone. There exists a Hamiltonian, providing oscillations around the stationary point
which is a dynamical circle \cite{FeEb89,EbScTi99}.
\be
Y_0 = \frac{b }{a}; \qquad X_0 = \frac{Q_0}{b} - \frac{c}{d}.
\ee
Note that this ecological system is a special case of a more general dynamical system studied by these authors
in earlier work
\cite{FeEb89,FeEb78}:
\be
{\dot X} = P + F(X) Y -  A_1 X - C Y; \qquad {\dot Y} = A_0 - F(X) Y
\label{Lotka2}
\ee
We  come back to eq.(\ref{Lotka1}) by assuming $F(X) = c + a X, A_1 = b, A_0 = Q_0, C = c$. Note that by including quadratic terms into $F(X)$, the passive system can be converted into an active one \cite{FeEb89,FeEb77,FeEb78}.\\

The definition of an external load is not trivial, since the introduction of the analogy of a "force" or "voltage"
makes difficulties. We restrict therefore our study to the case that the load is introduced by an unsymmetrical loss of the predator
species which acts only for increasing quantity of predators. As a model how this could work we may assume that with increasing
predator population the mortality increases due to competing mechanisms which generate an "ecological pressure"
against an increase of the predator population
\be
b \rightarrow b + b_1 > b \qquad if \qquad {\dot X} > 0.
\ee
Then the strictly Hamiltonian character is lost due to breaking the symmetry of ${\dot X}$, and the predator equation assumes the form
\begin{equation}
\label{Lotka3 }
  {\dot X} = a X Y - b  X  - b_1 X
\end{equation}
with the stationary solution under load
\be
X_1 = \frac{Q_0}{ b + b_1} - {c}{a}.
\ee
The additional pressure is $b_1$ and the population $X_1$ is decreasing under pressure.

\subsection{Efficiency of energy conversion against an ecological pressure}

\begin{figure}[htbp]
 \begin{center}
 \includegraphics[width=6cm,height=6cm]{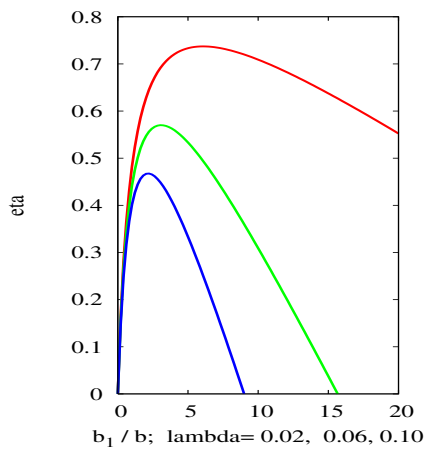}
  \caption{Lotka - Volterra system under ecological pressure: Efficiency $\eta$ vs. pressure - load parameter $b_1/b$ for different internal eco - parameters $\lambda = 0.02, 0.06, 0.10$. First the efficiency increases with the relative ecological pressure, reaches then a maximum around $b_1 / b \approx 3 - 6$ and goes then to zero. This means,
  the eco - system breaks down beyond some critical ecological pressure. }
 \end{center}
  \label{effieco}
 \end{figure}

Based on our simple model we can  calculate the efficiency of the transformation of energy against
an ecological pressure, just by mapping the present model to the previous one.
With ecological pressure we mean here an external action against the growth of a species.
For a Lotka - Volterra system of given type the pressure is  defined by an extra  loss and the additional effort to overcome this extra loss, providing the response of the predator population $b_1 {\dot X}_1$. Note that this term has the same dimension as the
energy transfer term $a X Y$. This way we find for the efficiency
\be
\eta = \frac{b_1 {\it X}_1}{Q_0} = b_1 \Big(\frac{1}{b + b_1} - \frac{c}{a Q_0}\Big) = \frac{b_1}{b} \Big(\frac{1}{1  + b_1 / b} - \lambda \Big)
\ee
with
\be
\lambda = \frac{b c}{a Q_0}
\ee
This entirely analytical theory tells us that the system works as converter of energy
only for $\lambda < 1$ and breaks down ($\eta \rightarrow 0$) if the parameter $R_1 / R$ is larger than $1 / \lambda - 1$.
The efficiency which we defined has a maximum  at
\be
\eta_{opt} =  \sqrt{\lambda^{-1} - 1}.
\ee
The value of the maximum is
\be
\eta_{max} =  \frac{\sqrt{\lambda^{-1} - 1} + \lambda}{1 - \sqrt{\lambda^{-1} - 1} }
\ee
and may be sufficiently near one, if only the $\lambda$ - parameter is sufficiently small (see Fig. \ref{effieco}).
We calculated in this section the efficiency of a model ecological system, defined as the relative part of the input energy which is used for the response against an external ecological pressure. Similar as the active circuit, the
active Lotka - Volterra system increases first its efficiency with the load, reaches then a maximum and breaks suddenly down beyond a critical ecological pressure.

\section{Conclusion}
This work is motivated by the first papers of Ulrike Feudel and intends to give some survey of the continuing discussion about increasing and optimizing
the efficiency of the conversion of available chemical
energy into higher forms of energy like mechanical, electrical or ecological energy.
Analyzing several examples, without claiming for any completeness, we arrive at the statements:\\
1. The conversion of chemical energy to higher energy forms by isothermal processes is evidently not subject to "fundamental" upper bounds, like the Carnot upper bounds for the conversion of thermal to higher energy forms. As we have shown by two examples, efficiencies can be higher than 50 percent. Such a high efficiency needs as a rule the exploitation of nonlinear effects and an optimization. This requires a
complete qualitative analysis of the properties of the nonlinear dynamical system.\\
2. The problem of optimization of efficiencies is connected with nonlinear dynamical systems with many system parameters.
For such a dynamical multi - parameter problem, the purely empirical search based on "trial and error" is like a "search in the fog"
and is at the end not a promising strategy. This leads to the request for more advanced search strategies and in particular for strategies which may be formulated and investigated fully analytically \cite{PPSN}.\\
In order to give a solvable example going beyond the early approach by Feudel et al. \cite{FeuFeEb84} we analyze one quite simple model
of converting chemical into electrical energy, a caricature of a fuel cell, and secondly a model of energy conversion
in an ecological Lotka - Volterra schema of one active
player and one passive (chemical) player.\\
We studied in the present work the efficiency of these two (mathematically closely related) models of electrical and ecological circuits with three dynamical components converting chemical
into useful (electrical or ecological) energy. We showed that these model systems allow complete analytical solutions
which may be used in particular for the optimization of efficiency. The model shows also some interesting properties which qualitatively reproduce the commonly expected behavior against load,
which also was found in several recent experiments \cite{Toyabe13}. Typically the efficiency increases first with the load, reaches then an maximum near to an optimal load and breaks down finally. High values above 50 percent may be reached only in a narrow regime of optimal load, for parameters which provide low internal dissipation and
allow an effective use of the nonlinear factors in the dynamics.

Acknowledgement: The authors thank E. Gudowska-Nowak, Yu.M. Romanovsky, L. Schimansky - Geier, and F. Schweitzer for discussions and valuable advice.

\end{document}